\documentclass[12pt]{iopart}
\usepackage[dvips]{graphicx}
\begin{document}
\title{Inhomogeneous phase separation and domain wall dynamics in the orthorhombically distorted La$_{0.87}$MnO$_x$}
\author{K. De, S. Majumdar, S. Giri}

\address{Department of Solid State Physics, Indian Association for the Cultivation of Science, Jadavpur, Kolkata 700 032, India}
\ead{sspsg2@iacs.res.in (S Giri)}
\begin{abstract}
We observe the interesting magnetization and resistivity results in the orthorhombically distorted self doped manganite, La$_{0.87}$MnO$_x$. A significant irreversibility between zero-field-cooled and field-cooled   resistivities as well as  magnetization data is clearly observed up to a temperature which is well above the both paramagnetic to ferromagnetic ($T_c$) and metal to insulator transition ($T_{MI}$) temperatures.  In the ferromagnetically ordered and metallic states the magnetotransport results suggest the influence  of domain wall dynamics. Interesting scenario of memory effects through the domain wall dynamics is strikingly observed in the relaxation of resistivity in a limited temperature region. The characteristic features of magnetotransport behaviour are suggested duo to the co-existence of structurally driven inhomogeneous phase separation and domain wall dynamics.
\end{abstract}

\maketitle

\section{Introduction}
The investigation on the hole doped manganites still demands special attention because of the wide range of intriguing magnetic and transport properties \cite{dagotta, salamon}. Recently, the self doped manganites with composition La$_{1-\delta}$MnO$_x$ also exhibit that the structural, magnetic and transport properties are very similar to the hole doped manganites. As a result of La deficiency, the oxidation state of Mn changes partially from Mn$^{3+}$ to Mn$^{4+}$ in association with the spectacular changes in the physical properties like the hole doping in manganites. The systematic studies in La$_{1-\delta}$MnO$_x$ show that the structural, electronic transport and magnetic properties strongly depend on $\delta$ and $x$ \cite{brion,markovich,troyanchuk,joy,sankar}. The maximum La deficiency in the single phase perovskite has been noticed in La$_{0.87}$MnO$_x$, where physical properties are found to be highly sensitive to $x$. In case of most reduced sample a paramagnetic to orbitally ordered antiferromagnetic (AFM) transition is observed like stoichiometric LaMnO$_3$, where the orthorhombic phase is noticed at room temperature. The gradual transition from orbitally ordered orthorhombic to orbitally disorder ferromagnetic (FM) state has been observed with the increase of oxygen stoichiometry. With the further increase of oxygen stoichiometry a structural change from orbitally disordered orthorhombic phase to monoclinic phase is observed at room temperature, where the dynamic orbital correlations favour FM state. Recently, the examples of metastabilities in the self doped manganites have been reported based on low-field dc and ac susceptibility measurements, which are mainly suggested due to the domain wall dynamics in the system \cite{sankar,muroi1,muroi2}. Because of the interplay among transport, magnetic, orbital, and structural degrees of freedom, the self hole doped manganites attracts considerable attention, which has not yet focused widely like hole doped and charge ordered (CO) manganites.

\par
In this paper, we report on the magnetic and electrical transport properties of La$_{0.87}$MnO$_x$ under different conditions of applied magnetic field and thermal cycling. The results exhibit the evidence of meatstable states over a wide temperature range, which is found even at much higher temperature than the paramagnetic (PM) to FM and metal to insulator (MI) transitions. The metastabilities in the magnetotransport behaviour are found to be strongly influenced by the structurally driven inhomogeneous phase separation (PS) and domain wall dynamics in the FM ordered state. In addition, an interesting feature of memory effect is observed over a limited temperature range  in the dynamics of resistivity for the orthorhombically distorted oxygen deficient compound. 

\section{Experimental}
The polycrystalline sample with nominal composition La$_{0.87}$MnO$_x$ was prepared by a chemical route as described in the literature \cite{de1}. The final heat treatment was performed at 1200$^0$C for 12 h in air, where the samples were cooled down to room temperature by furnace cooling. A portion of the sample annealed in air was further annealed in presence of oxygen at 1000$^0$C for $\approx$ 6 h. Single phase of the sample was characterized by an X-ray diffractometer (Seifert XRD 3000P) using CuK$_{\alpha}$ radiation. The powder X-ray diffraction analysis confirms the single phase of the compound, which could be indexed by the orthorhombic structure ($Pnma$). In accordance with the reported results the crystal structure of the oxygen annealed sample was indexed by the monoclinic structure ($I$2/$a$) \cite{troyanchuk}. For the simplicity, we address the oxygen annealed sample as La$_{0.87}$MnO$_x$(O$_2$). The magnetoresistance (MR) was measured by the standard four-probe technique, fitted with an electromagnet and a commercial closed cycle refrigerator operating down to 10 K (Janis Research Inc.). The magnetization ($M$) was measured using a commercial superconducting quantum interference device (SQUID) magnetometer (MPMS, XL). In zero-field cooled (ZFC) measurements of $\rho$ and $M$, the sample was cooled down to the desired temperature at zero magnetic field and the measurements were performed in the heating cycle just after the application of applied magnetic field. For the field-cooled (FC) condition, the sample was cooled down to the desired temperature with magnetic field and the measurements were performed while heating.

\section{Experimental results}
\subsection{Magnetic properties}
Temperature dependence of $M$ measured at 1 kOe is shown in the top panel of figure 1, indicating a PM to FM transition ($T_c$) at 130 K, which was obtained from the minimum of $dM/dT$ against $T$ plot. The transition is not sharp like a typical PM to FM transition, which is rather spread over a wide temperature region. At low temperature the magnetization indicates a decreasing trend with a broad peak around 15 K, while the ZFC magnetization ($M_{ZFC}$) deviates from the FC magnetization ($M_{FC}$) around $\sim$ 85 K. The inset of the top panel of figure 1 exhibits the magnetization curve at 5 K, which does not saturate even at 50 kOe. The value of moment at 50 kOe is found to be 2.77 $\mu_B$/Mn, which is less than that of the effective moment, $\mu_{eff}$ = 5.91  $\mu_B$/Mn, obtained from the linearity of the inverse susceptibility with paramagnetic Curie temperature, $\theta$ $\approx$ 200 K. Note that the value of $\mu_{eff}$ is much larger than the spin-only value for S = 2, $\mu_{theo}$ = 4.90 $\mu_B$, suggesting the existence of FM clusters in accordance with the reported results \cite{brion,troyanchuk,de1}. The inverse susceptibility deviates from the Curie-Weiss behaviour around $\sim$ 240 K, which is much higher than $T_c$ as seen in the bottom panel of figure 1. The orthorhombic crystal structure at room temperature and the value of moment at 50 kOe and 5 K suggest the oxygen stoichiometry, $x \sim$ 2.87 in consistent with the reported results \cite{troyanchuk}. 

\par	
We also measured the temperature variation of $M$ at low dc magnetic field, which is shown in figure 2. The temperature dependence of $M$ at 30 and 40 Oe exhibits different characteristic features compared to the measurement at 1 kOe. In the inset of figure 2, the strong ZFC-FC effect of magnetization is shown measured at 30 Oe. The $M_{FC}$ curve surprisingly deviates from the $M_{ZFC}$ around $\sim$ 230 K, which is consistent with the deviation of the inverse susceptibility around same temperature region from the Curie-Weiss law. In case of low field measurement $T_c$ obtained from the minimum of the $dM/dT$ against $T$ plot was found to be close to the high field measurements. The external magnetic field does not practically change the phase transition at $T_c$. A peak in the $M_{ZFC}$ ($T_p$) is noticed around 97 and 90 K for the magnetic field at 30 and 40 Oe, respectively, indicating that $T_p$ shifts toward the low temperature with increasing magnetic field. In addition to $T_p$, an anomaly ($T_a$) is noticed in $M_{ZFC}$ around $\sim$ 65 and $\sim$ 48 K for 40 and 30 Oe, respectively. In contrast to the field dependence of $T_p$, $T_a$ is shifted toward the higher temperature with magnetic field. Note that similar results were also reported in $M_{ZFC}$, where the anomaly in the $M_{ZFC}$ was in fact, suggested due to the domain wall pinning effects \cite{sankar}.

\subsection{Transport properties}
Temperature dependence of $\rho$ for La$_{0.87}$MnO$_x$ is shown in the top panel of figure 3, which was measured in zero magnetic field and 2 kOe magnetic field under both ZFC and FC conditions. Temperature dependence of $\rho$ in zero field exhibits a peak at around 144 K ($T_{MI}$), which is found to be much larger than $T_c$. The nature of the broadened peak at 144 K looks different in contrast to the sharp peak at the temperature of metal-insulator transition for hole doped compounds. In between 35 K and 153 K the typical metallic behaviour ($d\rho/dT > 0$) is noticed. The small increase of $\rho$ is clearly observed with a minimum in $\rho$ at 35 K, which is more pronounced in the temperature dependence of low-field MR. The effect of grain boundary contribution to the $\rho$ might have dominant role to the low-field MR for the polycrystalline sample. The effect of MR is noticed below $\sim$ 230 K, while the maximum of MR is observed near $T_c$, as seen in figure 3. 
The bottom panel of figure 3 shows the plot of $d\rho$/$dT$ against $T$ in the selected temperature region under different conditions of applied magnetic field. An anomaly is consistently noticed around 56 K in zero magnetic field, which does not change noticeably under ZFC condition. The anomaly is slightly shifted to 55 K under FC condition in accordance with the effect of external magnetic field dependence of $T_a$ observed in the low-field magnetization study (figure 2). It is interesting to point out that the present resistivity result is sensitive to the effect of domain wall movement even at zero magnetic field, if we believe the anomaly noticed in the magnetization study is ascribed to the effect of domain wall pinning \cite{sankar}.

\par
The bifurcation of ZFC and FC resistivities is clearly observed below $\sim$ 190 K (top panel of figure 3), which is  noticed at much higher than $T_c$ and $T_{MI}$. The values of $T_{MI}$ are found to be 149 and 153 K under ZFC and FC conditions, respectively. The bifurcation of ZFC and FC data has commonly been found in the temperature dependence of $M$, either due to  metastability  from the glassy component or from domain wall pinning \cite{mydosh}. The evidence of ZFC-FC effect in $\rho$ is rather, a rare example among transition metal oxides, which has recently been reported even in few manganites \cite{dho,smol}. A strong ZFC-FC effect in $\rho$ was observed in the phase separated (La$_{0.8}$Gd$_{0.2}$)$_{1.4}$Sr$_{1.6}$Mn$_2$O$_7$ below spin glass (SG) temperature ($T_{SG}$), where glassy behaviour was ascribed to the competing magnetic interactions between FM and AFM states \cite{dho}. The other example of ZFC-FC effect in $\rho$ was observed in CO compounds, where the CO state was destroyed by the application of strong magnetic field, resulting in the different $T$ dependence between FC and ZFC resistivities below the charge ordering temperature ($T_{CO}$) \cite{smol}. In the present case, the origin of FC effect in $\rho$ is different from the above cases rather, it shows another kind of example of FC effect in $\rho$ in the orthorhombically distorted the self doped manganites. In order to understand the origin, we incorporated the above experimental protocol for La$_{0.87}$MnO$_x$(O$_2$) as seen in figure 4. In contrast to the oxygen deficient sample, the bifurcation between ZFC and FC resistivities disappears. In place of broadened peak for oxygen deficient sample a sharp peak is observed at 260 K in association with a broad hump around 240 K. The features of the temperature dependence is similar to the reported results \cite{troyanchuk}, though the sharp peak in the present observation is 10 K larger than the sample with $x$ = 2.95. The temperature dependence of [$\rho$(0) - $\rho$(2 kOe)] clealy indicate a sharp peak at 258 K, where $T_c$ is expected, indicating $x$ $\geq$ 2.96 for La$_{0.87}$MnO$_x$(O$_2$).

\par
The weak thermal hysteresis in $\rho$ was strikingly observed in the metallic temperature region, when the measurement was performed in presence of magnetic field of 2 kOe under FC condition as seen in figure 5. The above scheme of measurements was performed both in absence and presence of magnetic field. The thermal hysteresis in $\rho$ was not observed in zero magnetic field. The irreversibility between cooling and heating curves starts around $\sim$ 155 K, which is well above $T_C$ and $T_{MI}$, and they join up around $\sim$ 55 K, where anomaly in the $d\rho/dT$ was noticed [bottom panel of figure 3] under FC condition. We performed several successive cycles of heating and cooling under the FC condition and the observed that thermal hysteresis was reproducible. Since, the measurements are performed using an electromagnet, the question of experimental artifact due to persistent current in the superconducting magnet can be ruled out. Note that Troyanchuk {\it et al}. \cite{troyanchuk} suggested the probability of first order phase transition at $T_c$, which was indicated indirectly by the analysis of magnetization curves around $T_c$ for the monoclinic phase of oxygen rich samples. We measured the resistivity for  monoclinic La$_{0.87}$MnO$_x$(O$_2$) under FC condition both in the heating and cooling cycles, which does not show any thermal hysteresis. However, the thermal hysteresis of resistivity was clearly noticed for the monoclinic  La$_{0.87}$Mn$_{0.98}$Fe$_{0.02}$O$_x$ under FC condition.  Since, the ionic radii of Fe$^{3+}$ and Mn$^{3+}$ are same, the inhomogeneity in the lattice is introduced by the random substitution of Fe$^{3+}$, which might be responsible for the thermal irreversibility in the resistivity under FC condition. Note that the existence of thermal hysteresis of $M$ as well as in $\rho$ were noticed for CO manganites, indicating the first order transition below $T_{CO}$, which have been observed even in absence of magnetic field \cite{smol}. In addition to the mixed valent manganites and cobalites, the existence of thermal hysteresis in $\rho$ and $M$, indicating the first order phase transition have also been noticed in the intermetallics \cite{roy, singh}.

\par
We measured the MR by varying low magnetic field, which is shown in figure 6. The low-field MR curves were measured at selected temperatures under ZFC and FC conditions as shown in top panel of figure 6. The non-linearity of the MR curves were noticed at 10, 55, 100 and 150 K, where non-linearity decreases with increasing temperature. Similar to the temperature dependent behaviour, the large difference between MR curves is clearly noticed under ZFC and FC conditions at 10 K, while the difference decreases at 55 K. At 100 and 150 K, the difference between ZFC and FC MR curves was not distinguishable. The values of MR, defined as [$\rho(0)$ - $\rho(H)$]/$\rho(0)$, are $\sim$ 23, 22, 19 and 22 \% at 5 kOe for 150, 100, 55 and 10 K, respectively under ZFC condition, where $\rho(0)$ and $\rho(H)$ are the resistivities in zero field and field. We did not observe any significant change of magnetization curves in the increasing and decreasing cycles of magnetic field at 100 and 150 K under ZFC and FC conditions. However, a clear evidence of hysteresis in the increasing and decreasing cycles was observed at 55 K under ZFC condition, which enhances considerably at 10 K as seen in bottom panel of figure 6. Note that Sankar et al. \cite{sankar} also reported the considerable increase of coercivity from the hysteresis of magnetization below $T_a$. Here, the anomaly in $d\rho$/$dT$ is observed around 55 K, where the hysteresis in MR curves are noticed for T $\leq$ 55 K in consistent with the magnetization results.

\par
The dynamics of $\rho$ was measured under different conditions. We found that the values of $\rho$ do not change with time in absence of external magnetic field in all the measured temperature range. An interesting scenario in the relaxation of $\rho$ was surprisingly observed only at a very limited temperature region. The sample was first cooled down to 55 K under FC condition with 2 kOe magnetic field from room temperature, waited for $t_w$ = 1 h and 1.5 h and then $\rho$ was measured with time just after switching off the field. The striking evidence of the sharp change of $\rho$ at $t_w$ in the relaxation of $\rho$ is found, as seen in figure 7. The relaxation path for $t_w$ = 1 h does not follow the relaxation path for $t_w$ = 1.5 h. The above features are similar to the aging in $M$, which have been commonly observed for SG compound, indicating the cooperative relaxation processes. In addition, the aging effect in $M$ is commonly observed below $T_{SG}$, where the relaxation rate of magnetization, $S$ = 1/$H$($dM$/$d$ln$t$), is found to be maximum around, $t$ = $t_w$ \cite{mydosh}. The evidence of such effect in $\rho$ is rather, a rare example even in mixed valent manganites. Recently, the sharp change in $\rho$ at $t_w$ was also reported by Wu {\it et al}. in cobalite, where the origin of such unusual features was not explained \cite{wu}. In accordance with this results, the sharp change in $\rho$ is directly observed in the $\rho$ {\it vs.} $t$ plot at $t$ = $t_w$, not in the relaxation rate of $\rho$. Note that the relaxation of $\rho$ was noticed only in the limited temperature region around anomaly in $d\rho$/$dT$ [bottom panel of figure 3], where the relaxation in magnetization was also reported by Sankar {\it et al.}, suggesting the effect of domain wall dynamics \cite{sankar}. We also measured the time dependence of $\rho$ for La$_{0.87}$MnO$_x$(O$_2$) both in presence and absence magnetic field at different temperatures, which do not show any noticeable time dependence in $\rho$. 

\section{Discussions}
The influence of domain wall dynamics has been reported in different manganites mainly based on the low-field magnetization studies. The evidence of domain wall dynamics has also been reported in La$_{1-x}$MnO$_3$, which was characterised by the temperature dependence and relaxation studies of dc and ac susceptibilities \cite{sankar,muroi1,muroi2}. The influence of domain wall dynamics on $\rho$ has also the significant role in different FM materials viz., epitaxial thin film \cite{kent}, (Ga,Mn)As \cite{chiba}, nanocotacts of polycrystalline Fe$_3$O$_4$ \cite{vers}. The influence of domain wall movement on the resistivity behaviour of ferromagnetic manganite is not unlikely, because the transport properties are strongly dependent on the magnetic states in manganites. The effect of domain wall movement on the resistivity has been reported in cases of thin film \cite{wolfman}. However, the report on the electron scattering correlated to the domain wall dynamics, which is reflected in the resistivity behaviour, is very rare in the polycrystalline samples of manganites \cite{alv}. Here, we observe the effect of domain wall movement in the temperature, magnetic field and time dependence of $\rho$. An anomaly in $d\rho$/$dT$ is noticed around 55 K. The hysteresis of MR curves is clearly observed for T $\leq$ 55 K. In case of relaxation of $\rho$, the sample was first cooled to 40 K from room temperature with magnetic field of 2 kOe, waited for $t_w$ and then the relaxation was measured just after switching off the magnetic field. In this experimental condition the following features may be suggested that the system can memorize the tracks of domain wall movements and the effect of the switching off the magnetic field at $t_w$, which is reflected in the dynamics of $\rho$ by the discontinuous change at $t_w$. The above experimental results show an interesting feature of memory effects in the relaxation of $\rho$, which is ascribed to the domain wall dynamics. 

\par
The values of $T_p$ in the low-field magnetization shift toward lower temperature with increasing magnetic field. The paramagnetic to FM transition is spread over a wide temperature range. A large difference between the ZFC and FC magnetization is observed in addition to the slow decreasing trend with temperature up to the lowest measured temperature even for the measurement at 1 kOe. The magnetization curve at 5 K does not indicate saturating tendency even up to 50 kOe magnetic field. The above features are the necessary ingredients of SG features in the system. In fact, the competition between orbitally ordered AFM and orbitally disordered FM states may lead to the SG phase, which has been observed in different cases of manganites. Note that the positive values of $d\rho$/$dT$ below $T_c$ exhibits the metallic states, which does not fit well with SG state. The neutron diffraction results confirm the long range FM ordering below $T_c$ \cite{troyanchuk}, which rules out the possibility of real SG state at low temperature. In such a case, the explanation of the existence of domain wall movement below $T_c$ is rather justified than the glassy states.

\par
The crystal structure, magnetic and transport properties of the present self doped oxygen deficient compound exhibits FM metallic state at low temperature, where the characteristic features do not reflect the normal features of typical ferromagnet. The weak thermal hysteresis is noted below 155 K, which indicates the features of first order transition. The thermal hysteresis in $\rho$ does not appear in case of monoclinic phase of oxygen annealed sample. Note that the thermal hysteresis again appears convincingly in case of low Fe substitution with monoclinic structure, where the random substitution of Fe introduces the inhomogeneities into the lattice. Recently, the existence of local clustering and structural disorder were suggested out by Troyanchuk {\it et al}. to establish an intrinsic chemical and structural inhomogeneity in the orthorhombically distorted La$_{0.88}$MnO$_x$ (2.86 $ \leq x \leq 2.91$) \cite{troyanchuk}. However, they discussed such possibility based on the value of moment, which is less than the value of full spin arrangement in the ferromagnetically ordered state. The structural and magnetic properties of the present compound characterize the oxygen stoichiometry in within the above limit, where the thermal hysteresis in $\rho$ directly shows the evidence of structurally disorder induced phase separation in the oxygen deficient self doped manganites. In fact, the recent theoretical model\cite{dagotta} supported by the experimental observation in low hole doped La(Sr)MnO$_3$ also exhibits the existence of intrinsic structural inhomogeneity in manganites \cite{shibata}. Here, the MI transition at 144 K is broadened in contrast to the sharp MI transition observed for hole doped manganites and the oxygen annealed self doped, La$_{0.87}$MnO$_x$(O$_2$). The temperature dependences of $\rho$ and $M$ surprisingly exhibit the clear evidence of ZFC-FC effect much higher than $T_c$, which suggests the short range FM clusters are developed well above $T_c$. In addition, inverse of susceptibility deviates from Curie-Weiss behaviour around $\sim$ 240 K, which is much higher than $T_c$ at $\sim$ 130 K. The above results are consistent with the existence of inhomogeneous phase separation, which is originated partially due to the growth of short range FM clusters of different distribution of sizes above $T_c$. In addition to the orbitally disordered FM clusters, the orbitally ordered AFM phase may coexist, which might be responsible for the structural inhomogeneities. However, the detailed microscopic experiment viz., neutron diffraction is suggested to confirm the origin of the structural inhomogeneities.

\section{Conclusion}
The magnetization and resistivity results exhibit interesting features in the orthorhombically distorted La$_{0.87}$MnO$_x$. The strong ZFC-FC effects of magnetization and low-field MR in the temperature and magnetic field dependence are clearly noticed at much higher temperature than $T_c$, which is suggested due to the inhomogeneous phase separation. On the other hand, the characteristic features of magnetization and MR suggest the strong influence of domain wall dynamics in the ferromagnetically ordered state. Interesting scenario of memory effects through the domain wall dynamics is strikingly observed in the relaxation of $\rho$ at a limited temperature region. The magnetotransport behaviours of the oxygen deficient compound are suggested due to the coexistence of structurally driven inhomogeneous phase separation and domain wall dynamics.

\par
One of the authors (S.G.) wishes to thank DAE, India for the financial support. The magnetization data using SQUID magnetometer were measured under the scheme of Unit on Nanoscience and Technology of Department of Science and Technology at IACS, Kolkata, India.

\newpage

\begin{figure}
\centering
\caption{In the top panel temperature dependence of magnetization under zero-field cooled (open circle) and field-cooled (closed circle) conditions measured at 1 kOe. The inset shows the magnetization curve at 5 K. The bottom panel exhibits the inverse susceptibility. The solid line indicate the Curie-Weiss behaviour. The arrow indicates the  $T_c$.}
\end{figure}

\begin{figure}
\centering
\caption{Temperature dependence of magnetization measured at 30 and 40 Oe in zero-field cooled (ZFC) condition. The arrows indicate the anomalies below the broadened peak. The inset exhibits the field-cooled effect of magnetization, where the arrow indicates the temperature of bifurcation between ZFC and field-cooled magnetizations.}
\end{figure}

\begin{figure}
\centering
\caption{In the top panel temperature dependence of resistivity in zero magnetic field, zero-field cooled (ZFC) and field-cooled (FC) conditions with 2 kOe field and temperature dependence of magnetoresistance, $\rho$(0) - $\rho$($H$), where $\rho$(0) is the resistivity in zero field and $\rho$($H$) is the resistivity with 2 kOe field. In the bottom panel, temperature dependence of $d\rho$/$dT$ in a selective temperature region indicating the anomalies in zero magnetic field, ZFC and FC conditions.}
\end{figure}

\begin{figure}
\centering
\caption{Temperature dependence of resistivity in zero magnetic field, zero-field cooled (ZFC) and field-cooled (FC) conditions with 2 kOe field for La$_{0.87}$MnO$_x$(O$_2$). Temperature dependence of magnetoresistance, $\rho$(0) - $\rho$($H$), where $\rho$(0) is the resistivity in zero field and $\rho$($H$) is the resistivity with 2 kOe field.}
\end{figure} 

\begin{figure}
\centering
\caption{ Thermal hysteresis of resistivity under field-cooled condition in between 155 and 55 K, as indicated by the arrows. The other arrows indicate the mode of thermal cycling.}
\end{figure}

\begin{figure}
\centering
\caption{ In the top panel field dependence of resistivity, $\rho$($H$) scaled by the resistivity in zero field, $\rho$(0) under zero-field cooled (ZFC) and field-cooled (FC) conditions at 150, 100, 55 and 10 K. In the bottom panel field dependence of resistivity under ZFC condition in the increasing and decreasing cycles of low magnetic field at 10 and 55 K.}
\end{figure}

\begin{figure}
\centering
\caption{ Time dependence of resistivity measured in zero magnetic field at 55 K, while the sample was cooled down to 55 K under field-cooled condition of 2 kOe field, indicating the sharp change in resistivity around different $t_w$. Arrows indicate the time at $t_w$.}
\end{figure}
\newpage
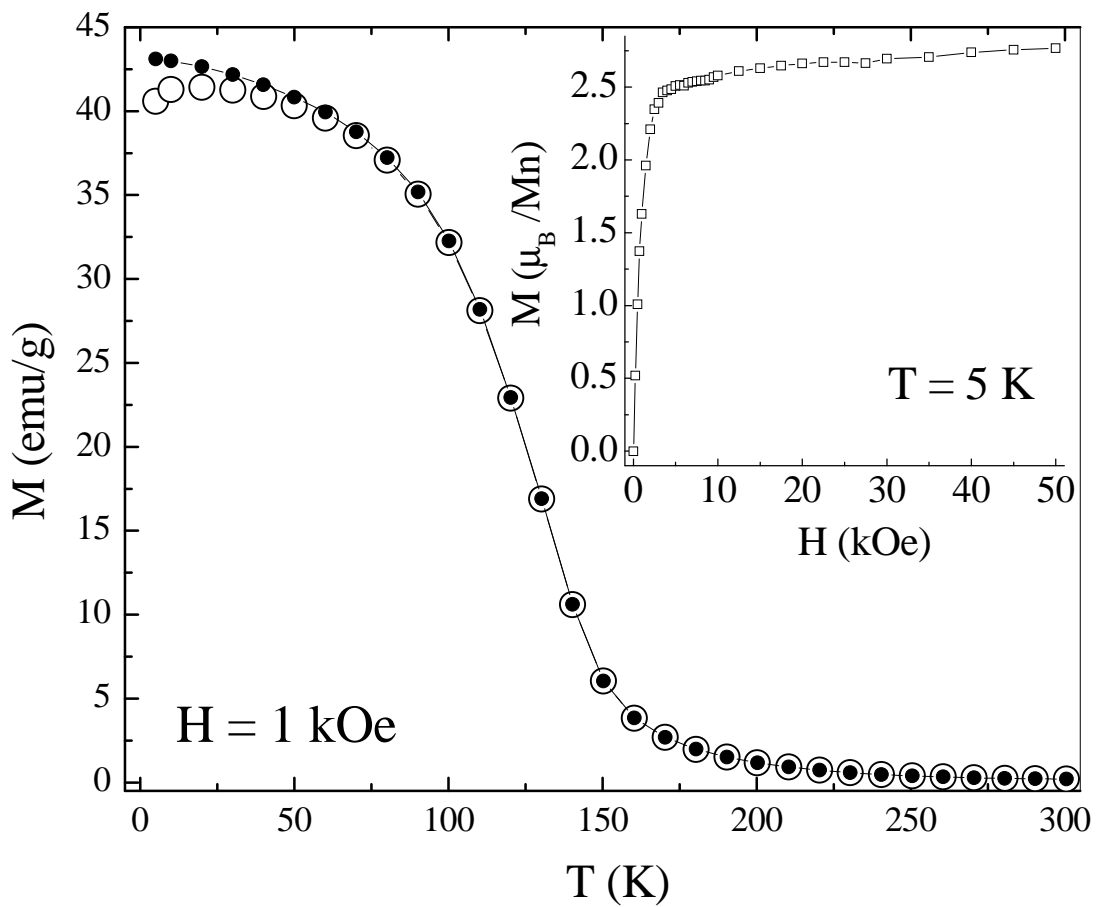
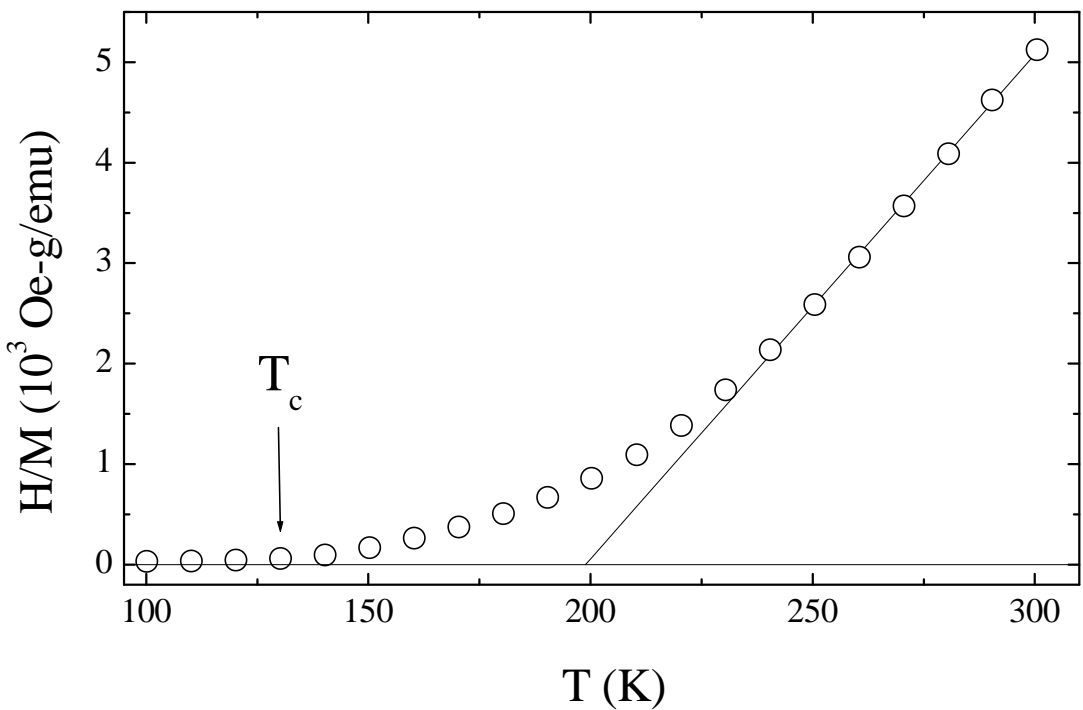
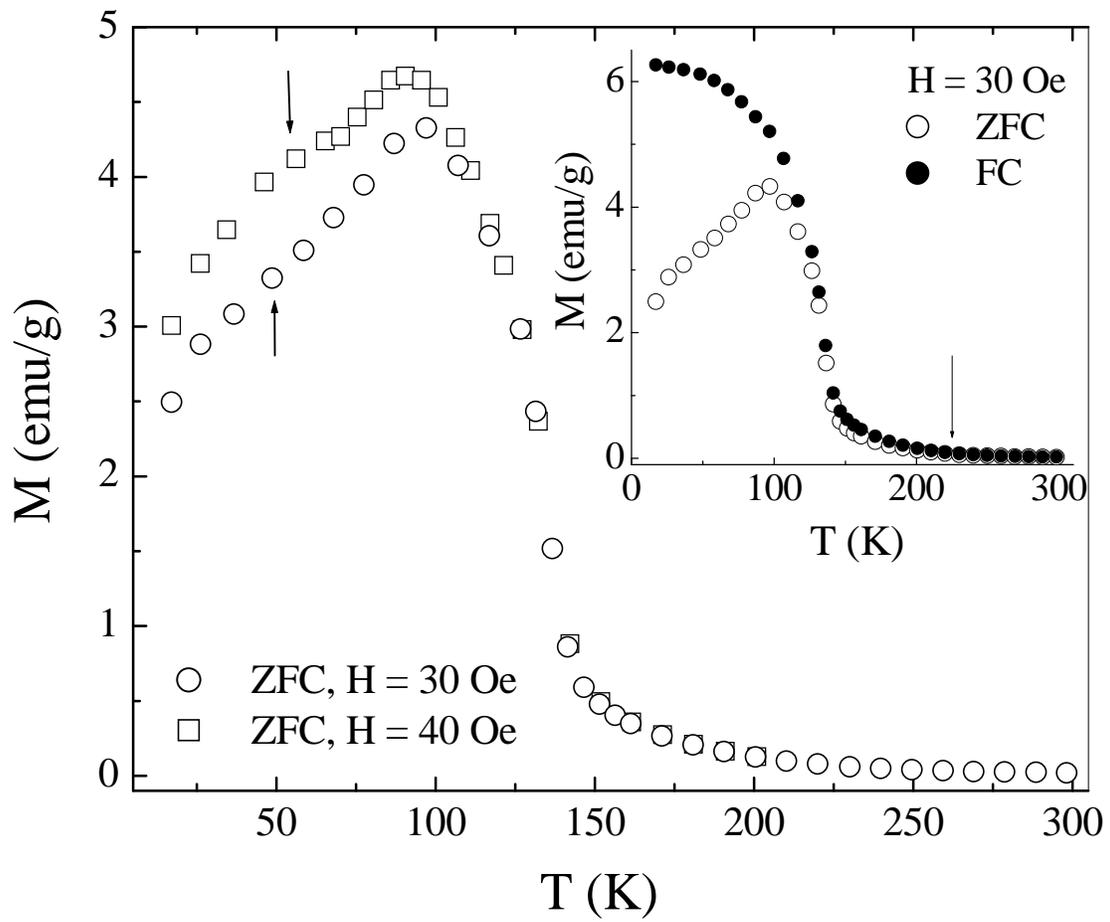
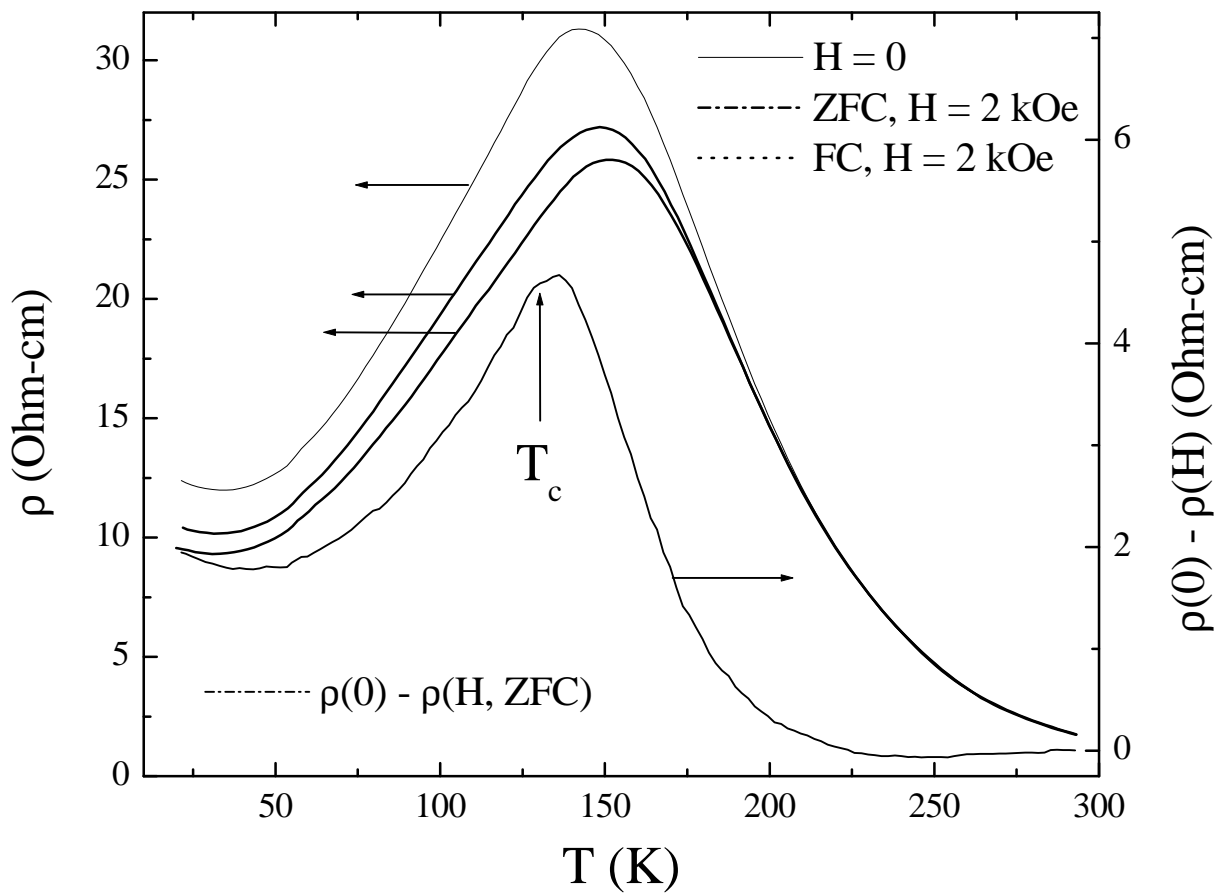
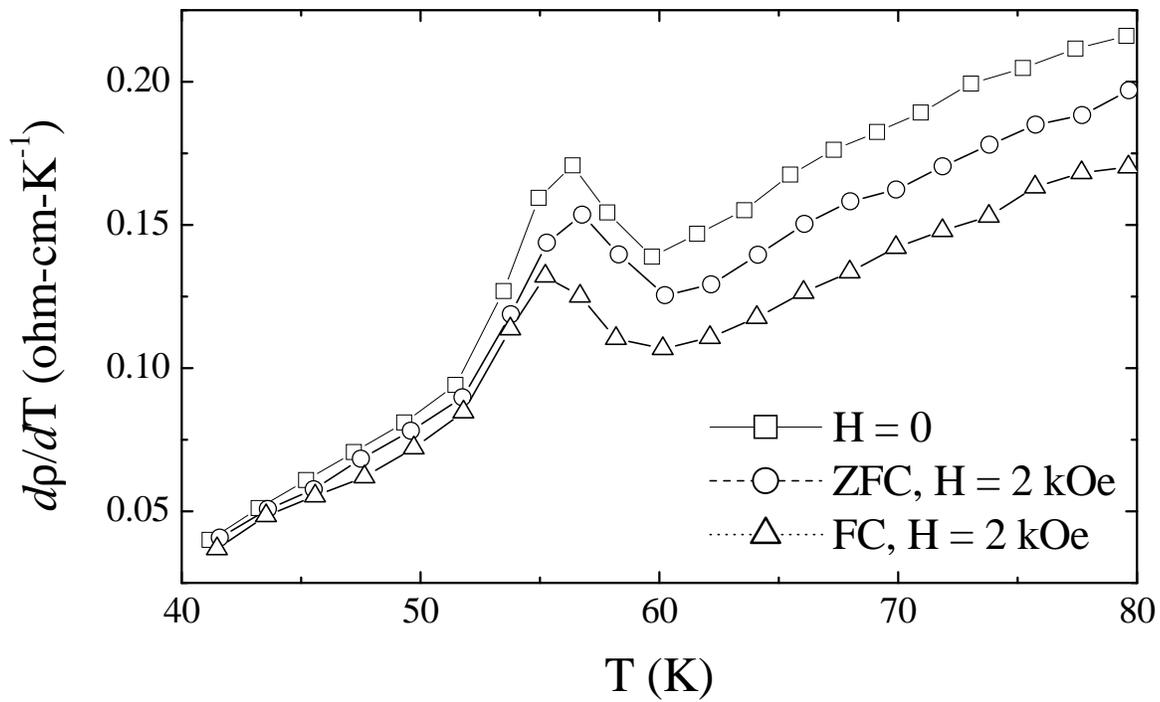
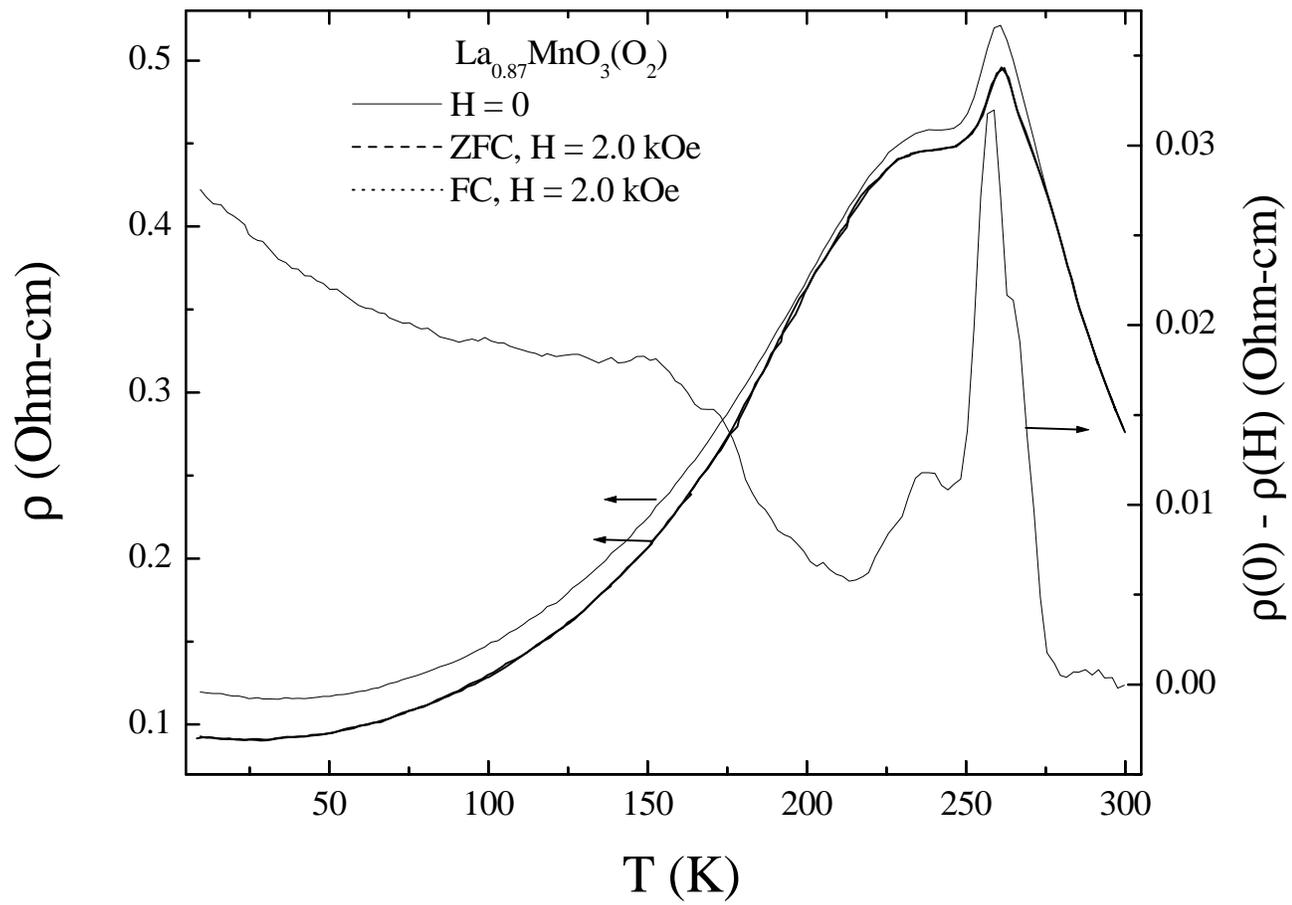
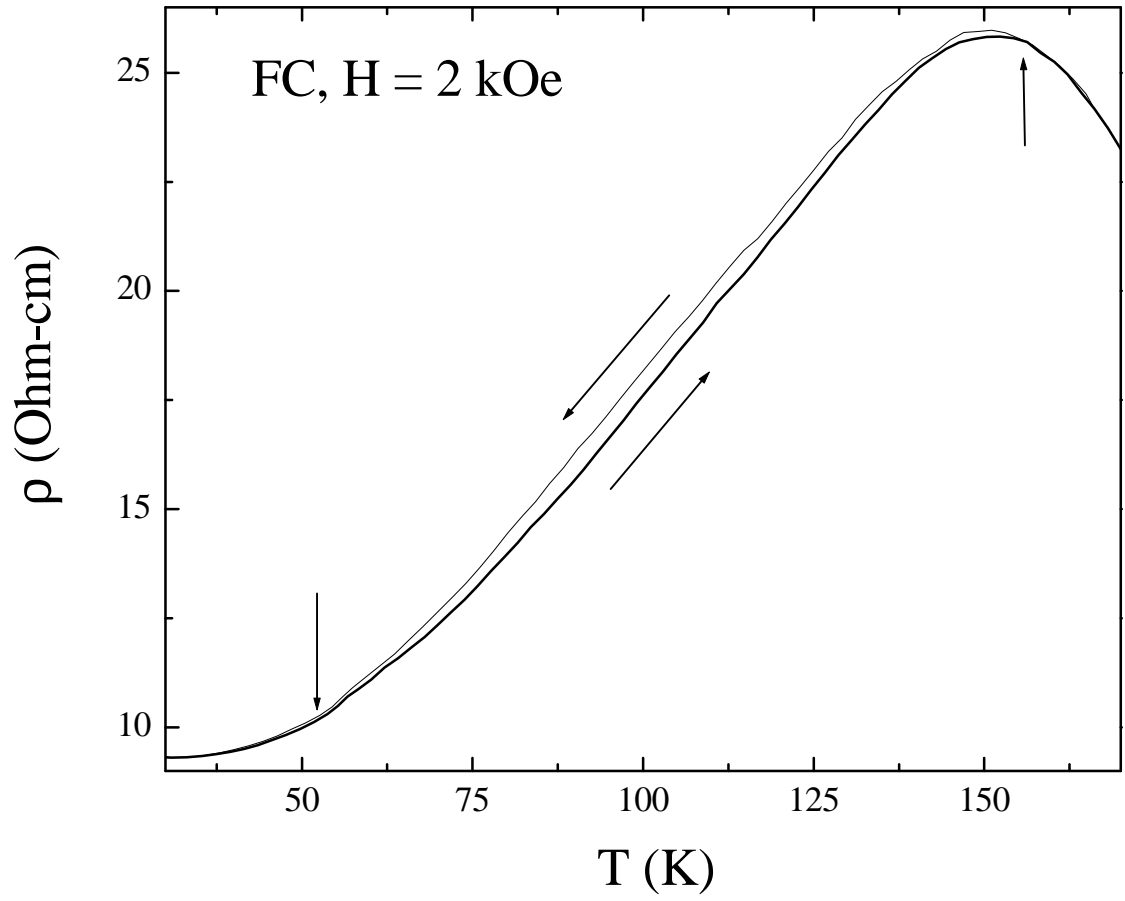
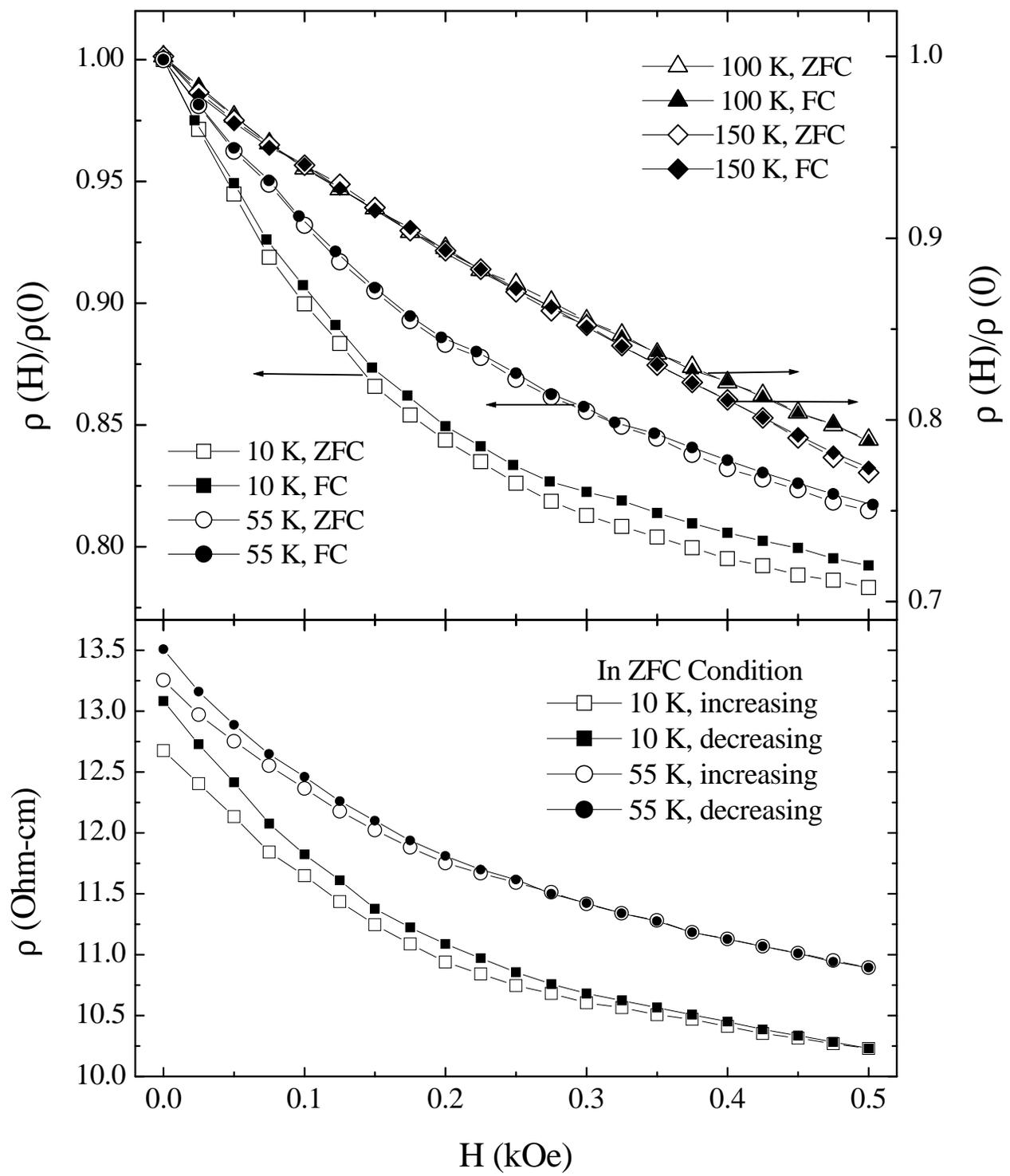
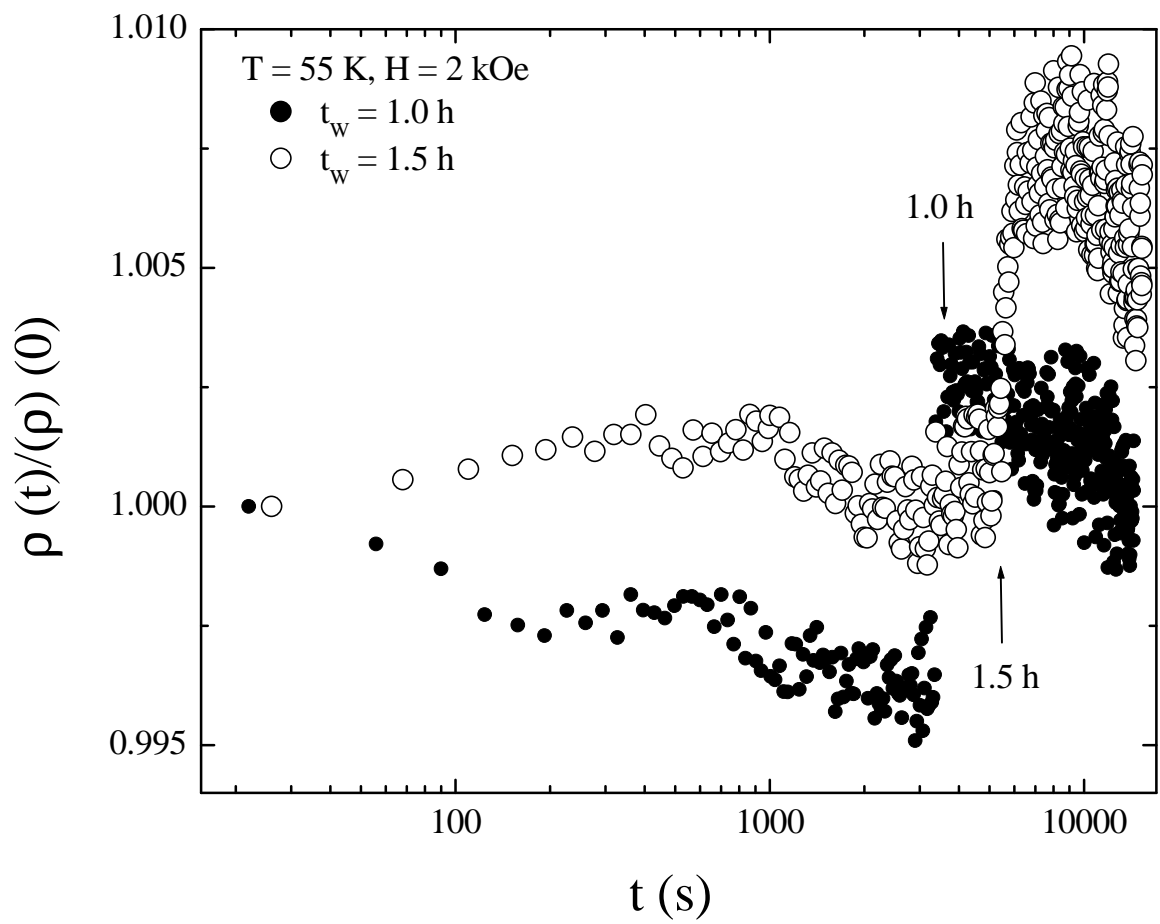
\end{document}